\documentclass[manuscript]{emulateapj}
\usepackage{amsmath}

\newcommand{\myemail}{rorysmith274@gmail.com}

\shorttitle{Tidal stripping}
\shortauthors{Smith et al.}

\begin{document}

\title{The Preferential Tidal Stripping of Dark Matter versus Stars in Galaxies}

\author{Rory Smith\altaffilmark{1}, Hoseung Choi\altaffilmark{1}, Jaehyun Lee\altaffilmark{2}, Jinsu Rhee\altaffilmark{1}, Ruben Sanchez-Janssen\altaffilmark{3}, \and Sukyoung K. Yi\altaffilmark{1} }  
\altaffiltext{1}{Yonsei University, Graduate School of Earth System Sciences-Astronomy-Atmospheric Sciences, Yonsei-ro 50, Seoul 120-749, Republic of Korea; \myemail}
\altaffiltext{2}{Korea Astronomy and Space Science Institute, Daejeon 305-348, Republic of Korea}
\altaffiltext{3}{UK Astronomy Technology Centre, Royal Observatory, Blackford Hill, Edinburgh EH9 3HJ, UK}

\begin{abstract}
Using high resolution hydrodynamical cosmological simulations, we conduct a comprehensive study of how tidal stripping removes dark matter and stars from galaxies. We find that dark matter is always stripped far more significantly than the stars -- galaxies that lose $\sim$80$\%$ of their dark matter, typically lose only 10$\%$ of their stars. This is because the dark matter halo is initially much more extended than the stars. As such, we find the stellar-to-halo size-ratio (measured using r$_{\rm{eff}}$/r$_{\rm{vir}}$) is a key parameter controlling the relative amounts of dark matter and stellar stripping. We use simple fitting formulae to measure the relation between the fraction of bound dark matter and fraction of bound stars. We measure a negligible dependence on cluster mass or galaxy mass. Therefore these formulae have general applicability in cosmological simulations, and are ideal to improve stellar stripping recipes in semi-analytical models, and/or to estimate the impact that tidal stripping would have on galaxies when only their halo mass evolution is known.
\end{abstract}

\keywords{methods: N-body simulations -- galaxies: halos -- galaxies: interactions}

\section{Introduction}
Galaxy clusters are the largest gravitationally bound structures to form within the large scale structure of the Universe. The gradient of the potential well of a galaxy cluster gives rise to strong gravitational accelerations, that drive a high velocity dispersion for galaxies residing within it. Nevertheless, it is not the net acceleration of a galaxy, but rather the difference in acceleration across the body of a galaxy, or tidal forces, that cause individual galaxies to suffer tidal mass loss. In \cite{Byrd1990}, the strength of the perturbation, that a galaxy experiences from the cluster potential, depends on its radius within the cluster to the inverse cubed. Thus the tidal forces are a smooth, decreasing function of radius, but the cluster potential is far more destructive near the cluster centre than in the cluster outskirts. The strength of the perturbation also depends on the physical size of the galaxy raised to the power of three. Therefore, at a fixed clustocentric radius, extended galaxies are much more perturbed. Tidal stripping from the cluster potential tends to preferentially affect the outer galaxy first (i.e. `outside-in' stripping). A simple approach to model this effect involves calculating the tidal radius of a galaxy (\citealp{BT1987}). Beyond this tidal radius, it is assumed that all material will be removed by external tides.

However, the cluster potential is not the sole source of tidal mass loss in clusters. Clusters are filled with other cluster member galaxies, with which tidal encounters can arise. Because the cluster galaxies typically have a high velocity dispersion ($\sim$1000~km/s), any galaxy-galaxy encounters tend to occur with high relative velocities. However, an individual galaxy may be subject to multiple, short-lived impulsive encounters. This process is known as `harassment' (\citealp{Moore1996}). The effects of such interactions can be approximated by the impulse approximation (\citealp{Gnedin1999}, \citealp{Gonzalez2005}). In the impulse approximation, the strength of the internal dynamical kicks that a galaxy receives from a high speed encounter depends on the impact parameter, encounter speed, and is linearly dependent on radius within the galaxy. Therefore, as with the potential of the cluster, tidal stripping from impulsive galaxy-galaxy encounters preferentially affects the outer galaxy first, resulting in outside-in stripping. 

The results of outside-in tidal stripping could potentially have important consequences for some galaxies. The halos and disks of galaxies may become truncated (\citealp{Smith2015}). In fact, preferential stripping of the more extended stellar body of a nucleated dwarf, leaving the central nucleus remaining, is one evolutionary route by which Ultra-Compact dwarfs may form (\citealp{Bekki2001}; \citealp{Pfeffer2013}; \citealp{Pfeffer2014}; \citealp{Ferrarese2016}). Also, disk galaxies in clusters may become increasingly bulge-dominated because their more extended disk component is preferentially tidally stripped compared to their bulge (\citealp{Aguerri2009}).

Observationally, it is often difficult to assess if a galaxy is clearly suffering harassment. The high speed nature of the tidal encounter means that the interacting galaxy may be long gone by the time we observe a harassed galaxy. Galaxies that undergo harassment often produce stellar streams, however the stellar streams are typically very low surface brightness (\citealp{Moore1996}; \citealp{Davies2005}; \citealp{Mastropietro2005}; \citealp{Smith2010a}). Therefore, understanding the effects of harassment on cluster galaxies has largely remained the realm of numerical simulations (e.g. \citealp{Moore1998}; \citealp{Gnedin2003b}). Simulations have revealed that a key parameter controlling the effectiveness of harassment is galaxy surface-brightness. Low surface brightness disk galaxies suffer much more significant disruption, morphological transformation, and stellar stripping than high surface brightness galaxies (\citealp{Gnedin2003a}; \citealp{Moore1999}). Simulations also find that the high speed tidal encounters can increase the mass loss beyond that from the main cluster potential alone by 10-50$\%$ (\citealp{Gnedin2003b}, \citealp{Knebe2006}, \citealp{Smith2013}). The strength of mass loss from harassment is very dependent on a galaxy's orbital parameters within the cluster (\citealp{Mastropietro2005}; \citealp{Smith2010a}; \citealp{Smith2013}; \citealp{Smith2015}). Galaxies with small pericentres, combined with low eccentricity suffer the highest mass loss as they spend the most time where the cluster tides are most harsh. However, even large eccentricity orbits can be destructive, if the pericentre is sufficiently small (\citealp{Smith2015}).

Tidal mass loss can also arise in groups of galaxies. Simulations show group preprocessing can influence group members, and that the inclination of a galaxy's disk to its orbital plane can influence the efficiency of stellar stripping (\citealp{Villalobos2012}). This same dependence has since been noted in cluster harassment simulations as well (\citealp{Bialas2015}). Indeed a significant fraction of galaxies, that may have suffered effects from the group environment, may be found in clusters by redshift zero (\citealp{Mihos2004}; \citealp{McGee2009}; \citealp{DeLucia2012}). The presence of kinematically decoupled cores in some cluster dwarf ellipticals could be direct evidence for the influence of the group environment (\cite{Toloba2014a}), as it is very difficult to form such features by harassment (\citealp{Gonzalez2005}). The high frequency of cluster galaxies with merger features may also provide evidence for preprocessing (\citealp{Sheen2012}; \citealp{Yi2013}).

The impact of stellar tidal stripping is not just important for galaxies themselves. The stars that are stripped in a cluster contribute to the build-up, and properties, of the Intra Cluster Light (ICL), and the Brightest Central Galaxy (BCG) (\citealp{DeLucia2007}; \citealp{Contini2014}). The N-body simulations of a cosmological cluster in \cite{Rudick2009} showed that as much as 40$\%$ of the ICL is formed from cold streams of stars that, themselves, are formed by tidal interactions with the BCG, or formed in galaxy interactions in groups before infalling into the cluster. In fact, the latter is more common at high redshift, before many clusters assemble, when the group environment was much more common.

The modelling of stellar tidal stripping may be very important in Semi-Analytical Models (SAMs). In SAMs, a dark matter only cosmological simulation is augmented with analytical recipes for how galaxies grow and evolve, including gas cooling, star formation, and stellar feedback. Among these analytical treatments for how the baryons should behave, it is necessary to consider how the stellar component of galaxies should respond to tidal stripping. A wide range of stellar tidal stripping recipes have previously been applied in the literature. In many cases, the stellar mass of a galaxy is not altered until the dark matter reaches some critical limit, and then all of the stars are instantaneously stripped. For example, in \cite{Somerville2008} this occurs when the halo is truncated to a single radial scalelength. In \cite{Guo2011}, it is when the host halo density at pericentre surpasses the density of the baryons of the satellite. An alternative critical limit is when the halo mass is reduced to the same mass as the total baryonic mass (e.g., \citealp{Guo2011}; \citealp{Lee2013}) . In alternative recipes, the stellar mass is decreased more smoothly. \cite{Contini2016} assumes that the stellar mass reduces exponentially, once the galaxy enters a host halo. Alternatively some SAMs only strip stars beyond the tidal radius of that galaxy (e.g., \citealp{Henriques2010}; \citealp{Kimm2011}; \citealp{Contini2014}). 

As we will demonstrate in Section \ref{stellarmassfuncsect}, the choice of tidal stripping recipe can impact on the shape of the stellar mass function. The stripping recipe could also impact the growth rate of the ICL and BCG (\citealp{DeLucia2007b}; \citealp{Lidman2013}), and potentially alter the stellar metallicity radial gradients in massive galaxies (\citealp{Contini2014}).

Numerical simulations that model both the halo and stellar mass of galaxies with live components can give some insights into how tidal stripping of stars occurs. \cite{Penarrubia2008} and \cite{Smith2013a} demonstrate that very high fractions of dark matter must be stripped before significant stellar stripping occurs. In \cite{Smith2013a}, for a large number of harassed dwarf galaxy models, it was found that typically 80-90$\%$ of the dark matter must be tidally stripped, just to remove 10$\%$ of the stars. However, the exact fraction of dark matter that must be stripped is found to depend on the size of the stellar disk of the galaxy, such that a smaller disk requires even stronger dark matter losses to be equally affected. Thus, in \cite{Smith2013a}, an attempt was made to link the efficiency of dark matter stripping to the efficiency of stellar stripping. In this study, we will attempt to extend on this previous analysis, by using fully cosmological simulations, and measuring the complete relation between bound dark matter fraction and bound stellar fraction, for galaxies suffering a full range of mass loss. In Section 2 we describe the numerical setup, in Section 3 we present our results, and in Section 4 there is a discussion and conclusion.

\section{Setup}
\subsection{The Hydrodynamic Cosmological Simulations}
\label{hydrosims}
We conduct zoom simulations of clusters of galaxies, using the adaptive mesh refinement code {\sc{ramses}} (\citealp{Teyssier2002}). Clusters are initially selected from a low resolution, dark-matter only, 200~Mpc/h cubic volume, using initial conditions generated by {\sc{mpgrafic}} (\citealp{Prunet2008}). In this study we assume a flat $\Lambda$CDM universe with a Hubble constant $H_{0}$=70.4~km/s/Mpc, a baryon density $\Omega_{b}$=0.0456, a total matter density $\Omega_{m}$=0.272, a dark energy density $\Omega_{\Lambda}$=0.728, a rms fluctuation amplitude at 8Mpc/h of $\sigma_{8}$=0.809, and spectral index $n$=0.963 consistent with Wilkinson Microwave Anisotropy Probe 7 (WMAP7) year cosmology \citep{Komatsu2011}.

Then, each selected cluster is zoomed, by first tracking back all particles within 3 virial radii of the cluster, then adding an additional four levels of nested initial conditions. Now, each cluster is resimulated, with a full baryonic physics treatment, and reaching a maximum spatial resolution of 760~pc/h. When 8 or more dark matter particles (or the equivalent mass in baryons) is present in a cell, it refines to the next level.

The baryonic physics treatment is deliberately chosen to match that used in the Horizon-AGN simulations (\citealp{Dubois2012}), and is also described in Choi et al. 2016 (in prep). In brief, we use the standard implementation of radiative cooling in Ramses. A look-up table is applied of the metallicity- and temperature-dependent, collisional equilibrium, radiative cooling functions from \cite{Sutherland1993}, for a mono-atomic gas of H and He, with a standard metal mixture. UV background heating is calculated using \cite{Haardt1996}, assuming z$_{\rm{reion}}$=10.4, consistent with WMAP7. We note that we do not use more recent UV background models (e.g. \citealp{Haardt2012}), however we do not expect this to have a significant influence on the main conclusions of this study because it is focused on tidal stripping. Star formation occurs above a critial density of 0.1~H~cm$^{-3}$, following a Kennicutt-schmidt law, with a star formation efficiency of 0.02. Supernova feedback is modelled as in \cite{Dubois2008}, where stars with mass greater than 10 solar masses explode as supernovae, 10~Myr after their formation, releasing 10$^{51}$~erg per 10~M$_\odot$ into ambient cells, in the form of kinetic and thermal energy. Gas cooling is suppressed for 10~Myr following a supernova, so as energy is efficiently deposited in the ambient gas. Formation of, and feedback from super massive black holes (SMBHs) is also considered following prescriptions in \cite{Dubois2012}. SMBHs tend to form where the density is peaked, and are treated as sink particles, which can accrete mass and merge. SMBH feedback can occur in `quasar' or `radio' mode, depending on whether the accretion rate surpasses the Eddington limit. 

The time step between each successive snapshots is approximately 75~Myr. Halos in each snapshot are identified using the AdaptaHOP method (\citealp{Aubert2004}), with a minimum number of dark matter particles of 64. We also rerun the halo finding code but on the stellar particle distribution in order to identify galaxies. For the stellar particles, we use the most massive sub-node method (\citealp{Tweed2009}) and, in practice, we detect galaxies down to $\sim$10$^8$~M$_\odot$/h. We limit our analysis to galaxies within the zoom region (within three virial radii of the cluster), so as to exclude poorly resolved galaxies. In each snapshot we match the stellar component of a galaxy to its halo by finding the closest galaxy to the centre of the halo. In cases where a satellite galaxy is mis-identified as the main galaxy, due to its stellar component temporarily passing close to the halo centre, we filter for rapid, shortlived changes in stellar mass. In order to track halos between snapshots, we follow the main progenitor galaxy up the merger tree. Galaxy merger trees are built with ConsistentTrees (\citealp{Behroozi2013}). All of our sample galaxies have a complete tree from redshift three until redshift zero.

Our galaxy sample is extracted from the hydrodynamical zoom simulations of three cosmological clusters. At redshift zero, Cluster 1 is a massive cluster with virial mass of 9.2$\times$10$^{14}$~M$_\odot$/h, and virial radius of 2.5~Mpc/h. Cluster 2 is significantly less massive, with a virial mass of 2.3$\times$10$^{14}$~M$_\odot$/h, and a virial radius of 1.6~Mpc/h. Cluster 3 is the least massive cluster, with a virial mass of 1.7$\times$10$^{14}$~M$_\odot$/h, and a virial radius of 1.5~Mpc/h. Cluster 1, 2, and 3 contribute 60, 27, and 13$\%$ of our final galaxy sample, respectively.

\subsection{Linking tidal mass loss of stars and dark matter in cosmological simulations}

\begin{figure}
\begin{center}
\includegraphics[width=8.5cm]{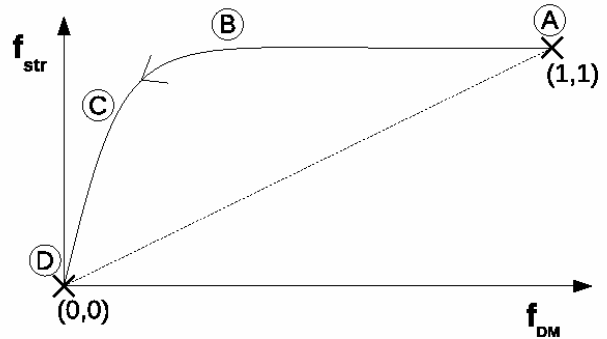}
\end{center}
\caption{Schematic of a toy galaxy undergoing tidal mass loss, and evolving along its dark matter-stellar mass fraction track. Position `A' marks the starting point of the galaxy, where it has all of its dark matter (f$_{\rm{DM}}$=1), and all of its stars (f$_{\rm{str}}$=1). The galaxy evolves to position `B', where it has lost more than half its dark matter, but none of the stars have yet been stripped. At position `C', the dark matter halo has been heavily stripped, and the stars begin to be stripped too. At position `D', the galaxy has been destroyed as all of the dark matter and stars have been stripped (f$_{\rm{DM}}$ and f$_{\rm{str}}$=0). The dashed line indicates the trajectory a galaxy would follow if it suffered equal dark matter and stellar mass loss.}
\label{sketchfig}
\end{figure}

In \cite{Smith2013a}, model early type dwarf galaxies were subjected to harassment. For each model galaxy, we measured the amount of dark matter that was stripped when exactly 10$\%$ of their stars were unbound. We found that the amount of dark matter was always very high, and a similar value was measured for all our model galaxies ($\sim$80-90$\%$). However, in this previous approach, we are only recording the link between the amount of dark matter and stars that are stripped at one specific moment, when exactly 10$\%$ of the stars have been stripped.

A more complete study can be made if we instead record the full relationship between the bound fraction of dark matter (f$_{\rm{DM}}$) and bound fraction of stars (f$_{\rm{str}}$), for galaxies that are undergoing tidal stripping. Note that, initially, before tidal stripping begins, a galaxy has a bound dark matter fraction, f$_{\rm{DM}}$=1. Then, as a galaxy suffers mass loss from tidal stripping of its dark matter halo, f$_{\rm{DM}}$ falls, approaching zero when the halo is almost entirely destroyed. We produce plots of f$_{\rm{str}}$ versus f$_{\rm{DM}}$, and each galaxy creates a track on such a plot, as it undergoes tidal mass loss. A schematic of such a plot is shown in Figure \ref{sketchfig}. We note that in \cite{Smith2013a}, our result represents a single point on such a curve. Therefore by considering the complete curve, we retain significantly more information on the galaxy's evolution while undergoing tidal stripping.

Our toy galaxy is initially at position A, before it suffers any tidal mass loss, and so its bound dark matter fraction and bound stellar mass fraction are both unity. However, as tidal stripping proceeds, the toy galaxy evolves along the track, towards position B. The initially horizontal motion of the track (between position A and B) indicates that dark matter is preferentially stripped, while no stars are stripped. Between position B and C, the dark matter halo has been heavily truncated and stellar stripping begins to occur, causing the turn-down in the track. Finally, the galaxy finishes at position D, where all of its dark matter and stars have been stripped, and the galaxy has been effectively destroyed. The dashed line is a one-to-one line. This is the trajectory a galaxy would follow if it suffered equal dark matter and stellar mass loss, at all times. Therefore, the fact that the track is always above the dotted line demonstrates that dark matter is always preferentially stripped in our toy model galaxy.

We note that when we refer to the total bound fraction of dark matter f$_{\rm{DM}}$, this fraction is derived from the total mass of the halo, which includes dark matter within the baryonic component of the galaxy, and dark matter at larger radii, out to the virial radius of the halo. Therefore we caution that our dark matter fractions should not be directly compared to observationally derived dark matter fractions. In most cases, the dynamics of a galaxy's baryons, such as a HI rotation curve, or stellar velocity dispersion, are used to derive their observational dark matter fractions. As such, the observations only probe the dark matter which exists within the radial extent of the baryons, which is typically just the inner dark matter halo. However, if the a galaxy's outer halo were stripped but its inner halo was largely unaffected, then our total bound dark matter fraction would be reduced, unlike the observationally derived quantity. Therefore the total bound dark matter fractions will often be more sensitive to tidal mass loss. In any case, a primary goal of our study is to improve stellar stripping recipes in SAMs, where calculating the total bound mass fraction of a halo can be accomplished directly from the N-body cosmological simulations on which the SAM is derived.

In \cite{Smith2013a}, `idealised' simulations were used that, although based on parameters from cosmological simulations, were not fully cosmological themselves. In this study we will use fully cosmological, hydrodynamical simulations which are summarised in the following section. Because galaxies first grew hierarchically in cosmological simulations, we measure when each galaxy's dark matter halo peaks in mass. At this instant, we assume each galaxy begins its journey along the f$_{\rm{DM}}$-f$_{\rm{str}}$ track, starting at position (1,1) (e.g. location A in Figure \ref{sketchfig}). As we wish to clearly understand the tidal stripping process, we exclude galaxies that undergo major mergers (i.e. more major than 1:5 mass ratio) since reaching their peak mass (occurring in about 20$\%$ of cases), as these can result in additional scatter in the motion of a galaxy along its track.

\begin{figure}
\begin{center}
\includegraphics[width=8.5cm]{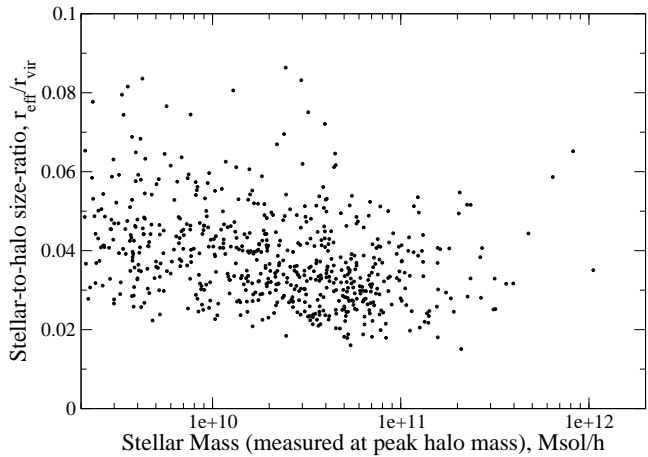}
\end{center}
\caption{Stellar-to-halo size-ratio (r$_{\rm{eff}}$/r$_{\rm{vir}}$) plotted against stellar mass. Both parameters are measured when each galaxy's halo reaches peak mass.}
\label{concmstrfig}
\end{figure}

Previous studies (\citealp{Penarrubia2008}; \citealp{Smith2013a}) have shown that if a galaxy has a smaller disk, more dark matter must be stripped to cause stellar stripping. Therefore, we divide our sample into three categories based on the relative size of their stellar component (measured using the effective radius r$_{\rm{eff}}$), compared to their dark matter halo (measured using the halo virial radius r$_{\rm{vir}}$). We form the stellar-to-halo size-ratio r$_{\rm{eff}}$/r$_{\rm{vir}}$, which is measured at the moment in which galaxies reach their peak halo mass. Galaxies with r$_{\rm{eff}}$/r$_{\rm{vir}}$$<$0.025 fall in our `concentrated' category, and galaxies with r$_{\rm{eff}}$/r$_{\rm{vir}}$$>$0.04 fall in our `extended' category. Galaxies that fall inbetween these limits are in the `intermediate category'. We choose these limits because, as we will show in Section \ref{deponconcsect}, these choices lead to a clear deviation in the response of the galaxies to tidal stripping between the subsamples. The concentrated, intermediate, and extended category make up 11, 50, and 39$\%$ of our final galaxy sample, respectively. In Figure \ref{concmstrfig}, we plot the stellar-to-halo size-ratio against the stellar mass of the galaxy, measured when the halo mass peaks. More massive galaxies tend to have slightly lower r$_{\rm{eff}}$/r$_{\rm{vir}}$, but the trend is not strong, and the spread is very broad at all stellar masses. Therefore the concentrated, intermediate, and extended galaxy subgroups each contain galaxies with a wide range of stellar masses.

In order to avoid undesirable numerical artifacts, we exclude all galaxies with r$_{\rm{eff}}$$<$2.5~kpc ($\sim$3 times the spatial resolution limit of our simulations). Additionally, the minimum detectable stellar mass of a galaxy is $\sim$1$\times$10$^8$~M$_\odot$/h. Therefore we exclude galaxies whose stellar mass is $<$2$\times$10$^9$~M$_\odot$/h when their halo mass peaks. This ensures we can measure f$_{\rm{str}}$ for all galaxies down to 0.05 or below. There is no imposed upper stellar mass limit. However the maximum stellar mass, measured when their halo mass peaks, is 1.1$\times$10$^{12}$~M$_\odot$/h in Cluster 1, 3.6$\times$10$^{11}$~M$_\odot$/h in Cluster 2, and 3.2$\times$10$^{11}$~M$_\odot$/h in Cluster 3.

In a final step, we separate our sample into two samples depending on the strength of their star formation. Galaxies whose f$_{\rm{str}}$ never rises above 1.15 (a maximum of a 15$\%$ increase in stellar mass) are placed in the `weakly star forming' sample. The remaining are considered to be forming stars more significantly. For most of the analysis of the paper we will focus only on the results of the weak star formation sample, which dominates the sample by number (82$\%$ of the sample). We exclude the strongly star forming galaxies from our final sample in order to understand the effects of tidal stripping of stars alone, while minimising the counter effect of new star formation. However, we expect that some galaxies may continue to form stars vigorously, even after reaching their peak halo mass. Therefore, an alternative recipe for tidal stripping of strongly star forming galaxies is presented in Section \ref{SFsection}. After applying previous cuts for major mergers, galaxy effective radius, minimum stellar mass, and now star formation, our final galaxy sample consists of 496 galaxies.

\begin{figure}
\includegraphics[width=8.5cm]{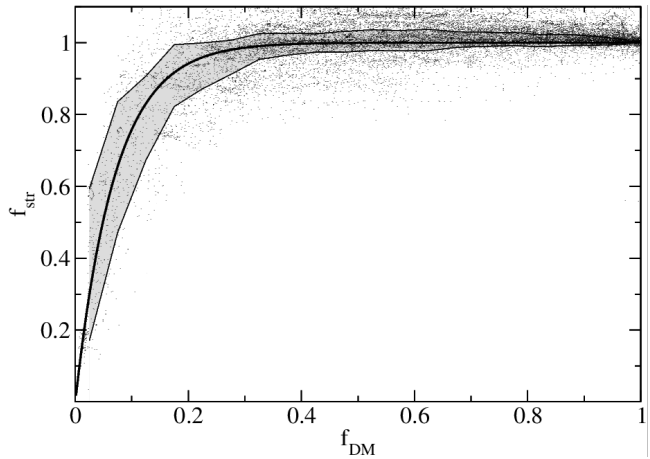}
\caption{A plot of the bound stellar fraction f$_{\rm{str}}$ versus the bound dark matter fraction f$_{\rm{DM}}$, for the final galaxy sample. The thick black central line is the fitted curve. Individual data points are shown as black dots. The grey shading indicates the first to the third quartile of the data points surrounding the fitted curve.}
\label{allinfig}
\end{figure}

We produce f$_{\rm{str}}$-f$_{\rm{DM}}$ plots, where each galaxy produces a single data point for every snapshot of the simulation since the galaxy reached peak halo mass. Because we only consider galaxies with weak or non-existent star formation, we find we can well fit the compilation of data points from all galaxies using a simple analytic form,

\begin{equation}
f_{\rm{str}}=1-\exp(-a_{strip} f_{\rm{DM}}){\rm{,}}
\end{equation}

\noindent
where a$_{\rm{strip}}$ is the unique exponential fitting parameter required to match the trend of the data points. In order to assess the degree of scatter about this line of best fit, we calculate the first and third quartile of the data points about the line of best fit.

\section{Results}
\subsection{The f$_{\rm{str}}$-f$_{\rm{DM}}$ diagram - no subcategories}
In Figure \ref{allinfig}, we show an f$_{\rm{str}}$-f$_{\rm{DM}}$ diagram that is based on our total final galaxy sample, without making any further subcategories for stellar-to-halo size-ratio. The bold line indicates the best fit to the data, and has the form
\begin{equation}
\label{allineqn}
f_{\rm{str}}=1-\exp(-14.20 f_{\rm{DM}}){\rm{.}}
\end{equation}

\begin{figure}
\includegraphics[width=8.5cm]{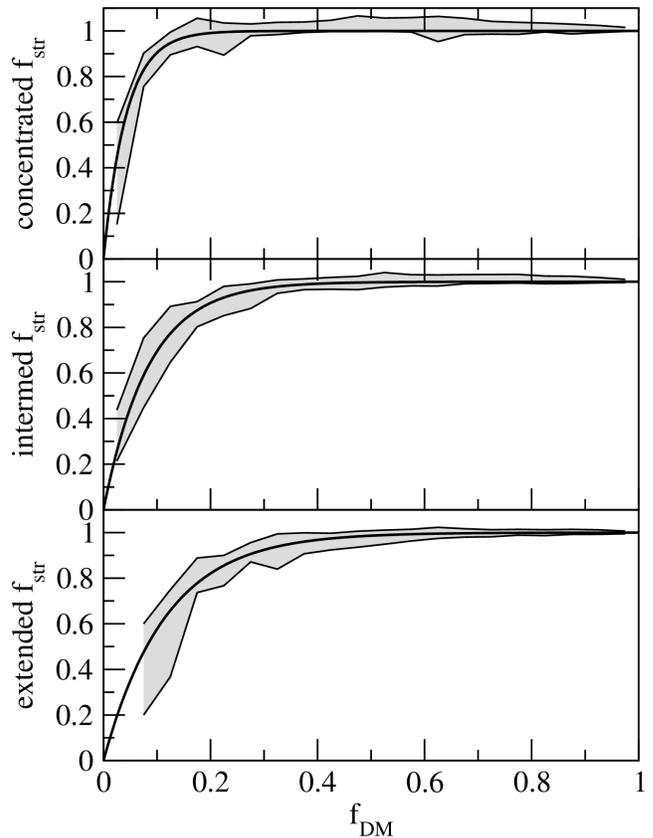}
\caption{The f$_{\rm{str}}$-f$_{\rm{DM}}$ plot for the galaxy sample, separated by how extended the galaxy stellar component is in comparison to the dark matter halo. The concentrated sample (top panel) has r$_{\rm{eff}}$/r$_{\rm{vir}}$$<$0.025. The intermediate sample (middle panel) has 0.025$<$r$_{\rm{eff}}$/r$_{\rm{vir}}$$<$0.04. The extended sample (lower panel) has r$_{\rm{eff}}$/r$_{\rm{vir}}$$>$0.04. The thick black central line is the fitted curve. The grey shading indicates the first to the third quartile of the data points surrounding the fitted curve.}
\label{starconcfig}
\end{figure}

The best fit line is steeply curved at small f$_{\rm{DM}}$, and remains in the upper left of the figure, indicating that the dark matter halos of the galaxies in the `no-subcategories' sample are always more susceptible to tidal stripping than the stellar component. Initially the dark matter fraction f$_{\rm{DM}}$ falls from 1 to $\sim$0.3, without any indication of stellar stripping. We find that when 10$\%$ of the stars are stripped, 84$\%$ of the dark matter has been stripped (compared to 85$\%$ for the standard model dwarf galaxy in \citealp{Smith2013a}). The upper and lower thin lines are the first and third quartile respectively. The distance between the first and third quartile is generally quite small, indicating that, in the absence of additional information on a galaxy, the bound dark matter fraction alone can provide a good first-order estimate of the fraction of stars that have been stripped. This suggests that Equation \ref{allineqn} could be useful for SAMs that have limited information on other galaxy properties, such as the size of the stellar component. However, as f$_{\rm{DM}}$ approaches zero (i.e. heavily tidally stripped), the spread about the best fit curve become increasingly large. In the next section, we will see that this is due to the range of size that the stellar component has, compared to the halo, in our sample of galaxies.

\subsection{Dependency on stellar-to-halo size-ratio}
\label{deponconcsect}
We now separate our sample into three subsamples, depending on their stellar-to-halo size-ratio. In Figure \ref{starconcfig}, our `concentrated' sample (r$_{\rm{eff}}$/r$_{\rm{vir}}$$<$0.025) is used in the top panel, our `intermediate' sample (0.25$<$r$_{\rm{eff}}$/r$_{\rm{vir}}$$<$0.04) is used in the middle panel, and our `extended' sample (r$_{\rm{eff}}$/r$_{\rm{vir}}$$>$0.04) is used in the lower panel.
 
In all three panels, the best fit curves deviate considerably from a one-to-one line, in a direction towards the upper-left corner of each panel, indicating that dark matter is always preferentially stripped from the galaxies. This indicates that the halo is significantly more extended than stars, even in the `extended' galaxy sample. However as we move from the `concentrated' to the `extended' sample, the curves increasingly approach the location of a one-to-one line. This indicates that if the sample has a smaller stellar-to-halo size-ratio, the galaxy must lose more dark matter to affect the stars. In other words, {\it{the more embedded the stars are within the halo, the more difficult they are to strip}} (\citealp{Penarrubia2008}).

Comparing with Figure \ref{allinfig}, the spread about the curve is generally decreased, in particular at small f$_{\rm{DM}}$. This is because the different trends seen in Figure \ref{starconcfig}, which differ most at small f$_{\rm{DM}}$, were being combined together in Figure \ref{allinfig}. This means that {\it{the most accurate prediction for stellar stripping is achieved if there is knowledge of the bound dark matter fraction, and the size of the stellar component}}. 

The best fit line for the concentrated sample (r$_{\rm{eff}}$/r$_{\rm{vir}}$$<$0.025) is
\begin{equation}
\label{conceqn}
f_{\rm{str}}=1-\exp(-23.94 f_{\rm{DM}}){\rm{,}}
\end{equation}

\noindent for the intermediate sample (0.025$<$r$_{\rm{eff}}$/r$_{\rm{vir}}$$<$0.04) is

\begin{equation}
\label{intermedeqn}
f_{\rm{str}}=1-\exp(-11.87 f_{\rm{DM}}){\rm{,}}
\end{equation}

\noindent and for the extended sample (r$_{\rm{eff}}$/r$_{\rm{vir}}$$>$0.04) is

\begin{equation}
\label{extendeqn}
f_{\rm{str}}=1-\exp(-8.60 f_{\rm{DM}}){\rm{.}}
\end{equation}

Fortunately, most SAMs (e.g. \citealp{Somerville1999}; \citealp{Cole2000}; \citealp{Hatton2003}; \citealp{Croton2006}; \citealp{DeLucia2007b}; \citealp{Somerville2008}; \citealp{Ricciardelli2010}; \citealp{Benson2010}; \citealp{Guo2011}; \citealp{Lee2013}; \citealp{Contini2014}; \citealp{Croton2016}, etc) already include prescriptions for the size of the stellar disk of their galaxies, based on the sharing of a fraction of a halo's angular momentum with its baryonic component (\citealp{Mo1998}). However, in the absence of information on the size of each galaxy's stellar component, the fit given in Equation \ref{allineqn} could be applied, albeit with less accurate results at small values of f$_{\rm{DM}}$.

\subsection{Negligible dependence on cluster mass or galaxy mass}
\begin{figure}
\includegraphics[width=8.5cm]{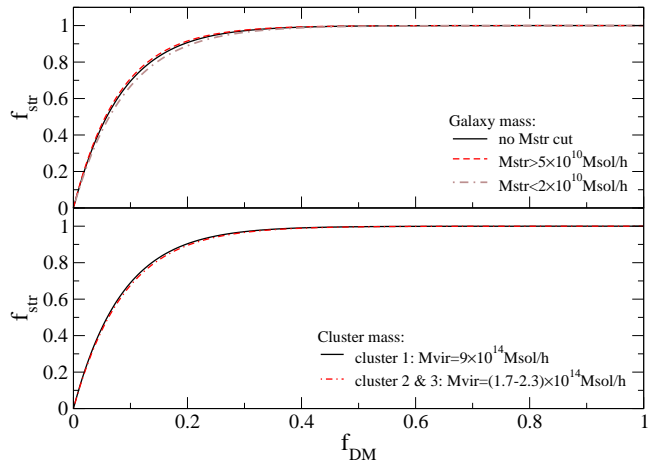}
\caption{Best fit curves for the dependency of the f$_{\rm{str}}$-f$_{\rm{DM}}$ curves on galaxy mass (top panel), and cluster mass (bottom panel). In both panels, the two separate curves lie closely on top of each other, illustrating the weak dependency of the curves on galaxy and cluster mass.}
\label{massfig}
\end{figure}

In the upper panel of Figure \ref{massfig}, we compare all of the galaxies in the `intermediate' sample (solid, black curve) to a `high stellar mass only' sample (containing only galaxies with a stellar mass $>$5$\times$10$^{10}$~M$_\odot$/h), and to a `low stellar mass only' sample (containing only galaxies with a stellar mass $<$2$\times$10$^{10}$~M$_\odot$/h). Previously we imposed a mass limit of 2$\times$10$^{9}$M$_\odot$/h, in order to limit numerical resolution effects. Thus the high stellar mass only sample has a cut at 25 times the value of the previous mass limit, and reduces the number of galaxies in our sample to 14$\%$ of its previous value. Despite the severity of the mass cut, the two curves are so close that it is difficult to separate them in the figure. Similarly, there is only a very minor difference between the curves, and the curve of the `low stellar mass only' sample. These results are important for two reasons. Firstly, it suggests that our results are not significantly affected by resolution effects, which would most likely have appeared in the lower mass galaxies. Secondly, the lack of a significant dependency on galaxy mass implies that the best fit lines (e.g. Equations \ref{allineqn}--\ref{extendeqn}) are universally applicable to galaxies of a wide range of sizes, which is very useful if they are to be applied to SAMs.
 
In the lower panel of Figure \ref{massfig}, we compare all of the galaxies in Cluster 1 (a massive cluster with a virial mass of 9$\times$10$^{14}$~M$_\odot$/h), to all the galaxies in Cluster 2 and Cluster 3 combined (lower mass clusters with a virial mass range of (1.7-2.3)$\times$10$^{14}$~M$_\odot$/h). Once again, the two best fit curves are difficult to distinguish, even though the cluster mass has changed by a factor of roughly four between the samples. This suggests that the behaviour of galaxies in response to the stripping of their dark matter is rather universal, independent of the mass of the system in which they reside. In fact, physically this makes sense. Even if a more massive cluster could cause stronger tidal stripping of galaxies, we see no obvious reason why such galaxies should deviate from the f$_{\rm{str}}$-f$_{\rm{DM}}$ relations that we have measured. It is logical that the {\it{relative}} tidal stripping of stars to dark matter would depend more sensitively on a galaxy's own properties, such as r$_{\rm{eff}}$/r$_{\rm{vir}}$, than the properties of the external potential. The lack of a dependency on cluster mass that we see once again supports the universal applicability of the best fit lines given in Equations \ref{allineqn}--\ref{extendeqn} to SAMs.

\subsection{A recipe for tidal stripping of star forming galaxies}
\label{SFsection}
We have so far only considered our main sample, which contains only `weakly star forming' galaxies. Therefore we now consider how to treat galaxies that continue to star form vigorously, after their halos reach peak mass. 

Previously, many SAMs assumed that a galaxy lost its gas content, and star formation was halted, as soon as a galaxy became a subhalo of another halo. It is likely that the moment at which a galaxy becomes a subhalo, is similar to the time when the halo mass reaches its peak value. Indeed, in our galaxy sample, only 18$\%$ of the galaxies increase their stellar mass by more than 15$\%$, since their halo's mass peaked. However, the assumption of a total halt in star formation has led to the `satellite over-quenching problem', where too many low mass satellite galaxies become quiescent, compared to observations (\citealp{Kimm2009}).

\begin{figure}
\includegraphics[height=5.0cm,width=8.5cm]{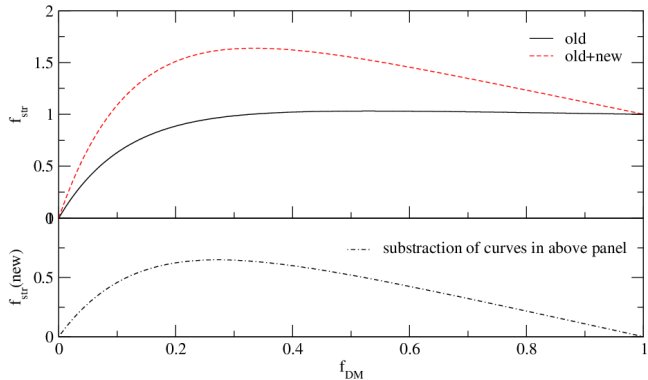}
\caption{In the upper panel we show the best fit curve for a sample of star forming galaxies, which fall in the intermediate category, whose total stellar bound fraction peaks between 1.5 and 1.75 (red dashed line, labelled `old+new'). The black curve shows the evolution of the stars formed before the halo mass peaked (black curve, labelled `old'). In the lower panel, the two curves are substracted (dot-dashed curve) to show the evolution of the fraction of new stars formed since the halo mass peaked.}
\label{SFfig}
\end{figure}

As a result, a number of authors have included recipes that permit galaxies to continue to star form, at least temporarily, after becoming satellites of a host galaxy. In some, a prescription for the removal of the hot gas content of a galaxy is employed (\citealp{Font2008}; \citealp{Kimm2011}; \citealp{Croton2016}). \cite{Tecce2010} also consider the gradual removal of the cold, atomic disk gas. Therefore, for these types of SAMs, it is necessary to consider a prescription for galaxies that may be undergoing tidal stripping, while simultaneously forming stars. 

In our star formation prescription, we split the total stellar mass of a galaxy into two components - `new' and `old'. We label all the stars formed prior to the moment when the halo mass peaks as `old' stars, and all of the stars formed since as `new', and calculate a bound stellar fraction for each component individually (f$_{\rm{str}}$(old) and f$_{\rm{str}}$(new)). 

We assume that the old stars are stripped in the same way as we have seen in the main/weakly star forming sample (i.e. f$_{\rm{str}}$(old) obeys Equations \ref{allineqn}--\ref{extendeqn}). We test this assumption, by measuring f$_{\rm{str}}$(old) directly from the cosmological simulation and find it is very reasonable.

To treat the new stars, we calculate f$_{\rm{str}}$(new) of a galaxy at each moment. This can increase if the galaxy continues to form stars. However, when the galaxy is heavily tidally stripped, and f$_{\rm{DM}}$ becomes small, we reduce the contribution of f$_{\rm{str}}$(new) to the galaxy's total bound stellar fraction f$_{\rm{str}}$. We choose that the contribution of the new stars is reduced by the same fractional decrease as f$_{\rm{str}}$(old) has decreased from unity. Mathematically, this can be expressed

\begin{equation}
\label{nustreqn}
f_{\rm{str}}=f_{\rm{str}}{\rm{(old)}}+f_{\rm{str}}{\rm{(new)}} \times f_{\rm{str}}{\rm{(old)}}
\end{equation}

In essence, we are assuming that the new stars are affected by tidal stripping in exactly the same way as the old stars are. In principle, this might not always be valid (for example, if new star formation should occur more centrally). However, our prescription could be easily modified to make the new stars more difficult to strip. For example, we could instead assume that new stars are not stripped until f$_{\rm{str}}$(old) reaches a critical value, f$_{\rm{crit}}$. Then Equation \ref{nustreqn} could become

\begin{equation*}
f_{\rm{str}}=\left\{
\begin{array}{@{}ll@{}}
f_{\rm{str}}{\rm{(old)}}+f_{\rm{str}}{\rm{(new)}}\text{,} & \text{if } f_{\rm{str}}{\rm{(old)}}>f_{\rm{crit}}\\
f_{\rm{str}}{\rm{(old)}}+f_{\rm{str}}{\rm{(new)}}\times \frac{f_{\rm{str}}{\rm{(old)}}}{f_{\rm{crit}}}\text{,} & \text{otherwise.}
\end{array}\right.
\end{equation*}   

Nevertheless, in practice we find that the new stars are stripped at a similar rate as the old stars, in our simulations. To test this, we first choose a sample of galaxies that were previously excluded from our main sample because they were forming stars too rapidly. We select galaxies with similar tracks on a f$_{\rm{str}}$-f$_{\rm{DM}}$ plot, by only choosing galaxies whose f$_{\rm{str}}$ has a peak value in the range 1.5 to 1.75 (we also test the range 1.25-1.5 but find the exact choice is not important to our conclusions). From this sample, we choose a subsample with 0.025$<$r$_{\rm{eff}}$/r$_{\rm{vir}}$$<$0.04 (i.e. they have a stellar-to-halo size-ratio that falls in the `intermediate' category). The red dashed curve in the upper panel of Figure \ref{SFfig} is a best fit line to these galaxies. The black curve in the upper panel is the best fit line for the `intermediate' sample (Equation \ref{intermedeqn}), and can be considered to trace the evolution of f$_{\rm{str}}$(old) of the sample. In the lower panel, we show the difference between the two curves in the upper panel, which is f$_{\rm{str}}$(new) for the sample. f$_{\rm{str}}$(new) initially grows as f$_{\rm{DM}}$ decreases from 1, because these galaxies are continuing to form stars. However we note that f$_{\rm{str}}$(new) begins to decrease from their peak value at nearly the same moment as f$_{\rm{str}}$(old) begins to decrease. In fact, f$_{\rm{str}}$(old) and f$_{\rm{str}}$(new) decline at a very similar rate, as f$_{\rm{DM}}$ becomes small. For example, f$_{\rm{str}}$(old) and f$_{\rm{str}}$(new) reach a value of 90$\%$ of their peak values, at roughly the same value of f$_{\rm{DM}}$ (f$_{\rm{DM}}$=0.19 and 0.16 respectively). Therefore, at least for the galaxies considered in our sample, the assumption that the new stars are equally affected by tidal stripping as the old stars appears valid, thereby supporting the use of Equation \ref{nustreqn}.

\subsection{Comparison to high resolution dwarf galaxy models}
\begin{figure}
\includegraphics[width=8.5cm]{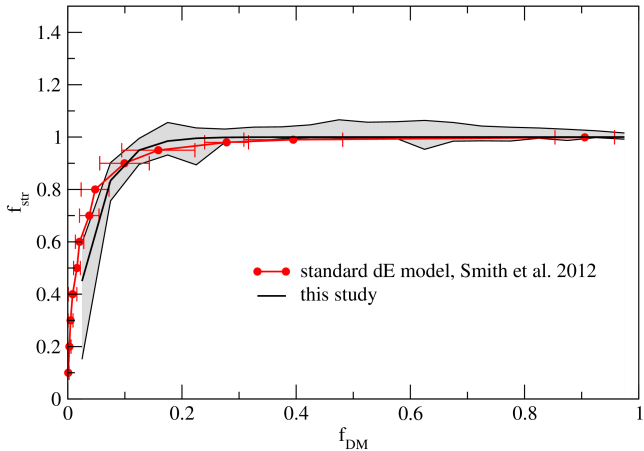}
\caption{The black curve is the `concentrated' category galaxies from this study. For comparison, the red symbols show the curve for the high resolution, early-type dwarf galaxies from the harassment simulations of Smith et al. 2013b. Error bars show the standard deviation of the multiple individual dwarfs used in that study.}
\label{2012compfig}
\end{figure}

By considering the complete evolution of a galaxy on a plot of f$_{\rm{str}}$ versus f$_{\rm{DM}}$, we can gather a more complete picture of how the galaxy responds to tidal stripping. Therefore we revisit the numerical simulations of \cite{Smith2013a}. We calculate how the standard early type dwarf model from our previous study evolves on a f$_{\rm{str}}$-f$_{\rm{DM}}$ plot. The results are shown in Figure \ref{2012compfig}. The red filled circles are the results from the 2013 study, where the error bars are the 1-sigma errors in the mean value. The standard early type dwarf model has r$_{\rm{eff}}$/r$_{\rm{vir}}$=0.013. This means its stellar-to-halo size-ratio clearly falls deep within in our `concentrated' category. Therefore we plot the `concentrated' best fit curve (Equation \ref{conceqn}) for comparison. 

This comparison is useful for two reasons. First of all, it provides a test of whether our prescription can be applied for galaxies down to the dwarf mass regime. It is important to test this as, in order to avoid numerical artifacts, we excluded galaxies with r$_{\rm{eff}}$$<$2.5~kpc from our sample, which effectively excluded dwarf galaxies from our sample. Secondly, the gravitational resolution of the early type dwarf simulations in \cite{Smith2013a} was only 100~pc -- roughly ten times better than the cosmological simulations. Therefore, it could also potentially identify if our new results were impacted by their more limited resolution. 

As demonstrated in Figure \ref{2012compfig}, we find that both our new results, and the early type dwarf model show similar behaviour. Both lose very large amounts of dark matter before stellar stripping becomes significant. In fact there is excellent agreement between the two studies for galaxies over a wide range of f$_{\rm{DM}}$ from 0.1 to 1.0. However for f$_{\rm{DM}}$$<$0.1, the early type dwarf model systematically sits above the best fit curve. However the offset is not substantial, and the early type dwarf models are found at approximately the position of the third quartile of the galaxies in this study. It is difficult to understand the true origin of this offset, as it could arise for multiple reasons. The significantly high gravitational resolution of the early type dwarf models could enable their self-gravity to be better resolved at their innermost radii, allowing them to better hold onto their stars. However, we also note that with r$_{\rm{eff}}$/r$_{\rm{vir}}$=0.013, the early type dwarf model is highly concentrated, and so might be expected to be slightly more robust to stellar stripping than the average galaxy in our `concentrated' sample. 


Nevertheless, we conclude that the broad agreement between the curves shown in Figure \ref{2012compfig} demonstrates that our recipes for stellar stripping are applicable in the dwarf regime, and suggest our results are not strongly altered by resolution effects. 

\subsubsection{Application to a SAM: the stellar mass function}
\label{stellarmassfuncsect}
We apply our new tidal stripping recipe (specifically Equation \ref{allineqn}) to the semi-analytical model, ySAM.  This SAM was developed by \citet{Lee2013}, and is based on a cosmological N-body volume simulation of structure formation, that was simulated using {\sc{GADGET-2}} \citep{Springel2005}. The cosmological parameters used match those in our hydrodynamical cosmological simulations (see Section \ref{hydrosims}). The periodic cube size of the volume is 200/h~Mpc on a side, with $1024^3$ collisionless particles. We generated a halo catalog by identifying substructures using SUBFIND \citep{Springel2001}. Then, halo merger trees were constructed from the halo catalog by a tree building algorithm described in \citet{Jung2014}. These merger trees were used as an input to the semi-analytic model. Further details of the baryonic physics prescriptions can be found in \citet{Lee2013}. 

\begin{figure}
\includegraphics[width=9.0cm]{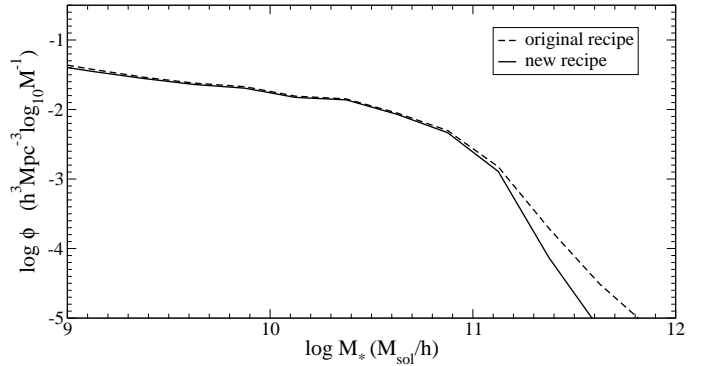}
\caption{Comparison of the stellar mass function of the galaxy population at redshift zero, produced by ySAM with the new tidal stripping recipe (solid line), compared to with the original tidal stripping recipe (dashed line).}
\label{SAMcompfig}
\end{figure}

In Figure \ref{SAMcompfig}, we compare the stellar mass function at redshift zero produced by ySAM, using the new tidal stripping recipe (solid line), compared to using the original tidal stripping recipe (dotted line). In the original tidal stripping recipe of ySAM there is no stellar stripping until the dark matter halo mass is equal to the baryonic mass of the galaxy (as also applied in \citealp{Guo2011}). Then, all of the stars are stripped instantly from the galaxy, and are added to the halo stars of the host galaxy.  The deviation between the two curves becomes clear above $\sim$10$^{11}$~M$_\odot$/h. The new recipe reduces the number of massive galaxies, and the offset increases with increasing stellar mass, thereby steepening the high mass end of the stellar mass function.

This offset likely arises because the most massive galaxies tend to accrete most of their stellar mass through mergers (\citealp{Lee2013}). Meanwhile these massive galaxies are preferentially found in a high density environment, such as a group or cluster, where tidal stripping of satellites is more likely to occur prior to merging. However, the tidal stripping of the satellites may be difficult to see in Figure \ref{SAMcompfig}, because the sample is drawn from a large cosmological volume, and so is dominated by low mass galaxies inhabiting low density environments. Therefore, we would expect even stronger effects, over a wider range of stellar mass, if we were to focus on the SAM results for high density environments only.

As is customary with SAMs, parameters that control baryonic physics recipes are tuned in order to match observed galaxy relations, such as the luminosity function. Therefore it is possible that some articifial over-tuning of parameters may have occurred, in order to compensate for inaccuracies in those earlier tidal stripping recipes that were employed. We briefly consider likely physical parameters that might have been affected in this way. As the original ySAM tidal stripping recipe creates a larger number of massive galaxies, the strength of AGN feedback may have previously been overestimated, compared to what is required with the new tidal stripping recipe. Altering the strength of the supernova feedback is a less desirable option, as this could impact on the stellar mass function at lower galaxy masses too. Alternatively, when a merger occurs, it is assumed that a fixed fraction (20$\%$) of the stellar mass of the galaxy is scattered into the stellar halo of the host galaxy (\citealp{Monaco2006}; \citealp{Murante2007}), instead of joining the main stellar mass of the galaxy. Therefore this fraction may have been set too high, compared to what is required with the new tidal stripping recipe. A number of other galaxy parameters, which are closely tied to a galaxy's stellar mass, are also likely to be influenced, including mass weighted age, metallicity, and disk-to-bulge ratio. We will explore these, and other consequences of our new recipe for SAM galaxy populations, in a future paper.

\section{Discussion and conclusions}
Using high resolution hydrodynamical cosmological simulations of three galaxy clusters, we have studied how the fraction of bound dark matter and stars is related for galaxies undergoing tidal stripping. We find that, in all galaxies, substantial quantities of dark matter must first be stripped, before stellar stripping can begin. Typically galaxies that lose $\sim$80$\%$ of their total dark matter, lose only 10$\%$ of their stars. We emphasise that we measure the total bound fraction of dark matter, which includes all the dark matter -- both inside and beyond the radii of the baryons. Therefore, our dark matter fractions are not directly comparable to observationally derived dark matter fractions, which typically can only probe the inner halo, where baryons are present. 

We find that the ease with which stars are stripped depends on a key parameter - the ratio of the effective radius of the stellar component, compared to the virial radius of the dark matter halo (r$_{\rm{eff}}$/r$_{\rm{vir}}$). We term this ratio the `stellar-to-halo size-ratio'. If the stellar component is more extended with respect to the halo, the stars are more easily stripped. With simple analytical fitting formulae (Equations \ref{allineqn}--\ref{extendeqn}), we quantify the link between bound dark matter fraction and stellar fraction. 

These fitting formulae could be applied to improve stellar stripping recipes in Semi Analytical Models (SAMs), or other numerical models of galaxies where only a live dark matter component is considered. With knowledge of only the bound dark matter fraction, a first order estimate of the evolution of the stellar bound fraction can be made using Equation \ref{allineqn}. However, if a galaxy's stellar-to-halo size-ratio is known, a more accurate prediction of the bound stellar fraction can be made using Equations \ref{conceqn}--\ref{extendeqn}. The improvement in accuracy is greatest when tidal stripping of the dark matter halo is very strong. We find negligible dependence on galaxy mass, and/or cluster mass, which suggests these equations can be applied universally, making them ideal for application in SAMs. We also provide a suggested recipe for the treatment of galaxies that continue to form stars, while suffering tidal stripping, in Section \ref{SFsection}.

The small scatter seen in the trends suggests that accurate predictions for the stellar stripping of a galaxy can be made, based on knowledge of the bound dark matter fraction, combined with a galaxy's stellar-to-halo size-ratio. It has been suggested that galaxy rotation could also be a factor dictating the efficiency of stellar stripping (e.g. \citealp{Donghia2009}; \citealp{Villalobos2012}; \citealp{Bialas2015}). Our cosmological simulations contain galaxies with a wide range of rotational properties (Choi et al. 2016 in prep.). Of course, it is possible that rapidly rotating galaxies might also have more extended stellar components. However, if rotation were important to our results, then the range of rotational vector with respect to their orbital plane should cause a wide range in efficiency of stellar stripping. Given the small spread that we measure about the trends, we conclude that rotation does not play as significant a role as stellar-to-halo size-ratio, at least in our simulations.

For some galaxy morphologies the effective radius, alone, might poorly encapsulate the mass distribution of the stars. One example is the case of a galaxy with a massive compact bulge, and an equally massive extended disk. In this scenario, calculating a separate stellar fraction for each of the two stellar components of the galaxy might, in principle, improve the accuracy of predicted stellar stripping. However, given the tightness of the trends, we note that this occurrence does not seem to arise frequently in our cosmological simulations. 

In summary, the equations provided in this study can be easily applied to SAMs, to improve existing stellar stripping recipes. In Section \ref{stellarmassfuncsect} we applied our recipe to ySAM (\citealp{Lee2013}), and found that it steepened the high mass end of the stellar mass function. We anticipate that the impact will be even stronger in higher density environments, such as groups and clusters. It will also have consequence for SAM predictions of the growth rate (\citealp{Lidman2013}), and metallicity gradients of brightest central galaxies, and intracluster light (\citealp{Contini2014}). We will fully explore the consequences of applying these improved recipes in SAMs in a future paper. In this paper we have focussed on the total dark matter fraction of the halos of galaxies, as this quantity can be easily measured in SAMs. However, in the future, it would also be interesting to calculate the fraction of dark matter contained within the baryons, as in this way the results could be compared more directly with the observations of real galaxies. As the dark matter halos are preferentially stripped from the outside inwards, we note that our bound dark matter fractions, which are measured for the total halo, represent lower limits for the equivalent quantity measured within the baryons.

\acknowledgments
RS acknowledges support from Brain Korea 21 Plus Program(21A20131500002) and the Doyak Grant(2014003730). SKY acknowledges support from the National Research Foundation of Korea (Doyak grant 2014003730). SKY, the head of the group, acted as a corresponding author. Numerical simulations were performed using the KISTI supercomputer, under the program of KSC-2014-G2-003. We are grateful to the referee for their constructive input.

\bibliographystyle{apj}
\bibliography{bibfile}

\clearpage

\end{document}